# Changes in single K$^+$ channel behavior through the lipid phase transition

Heiko M. Seeger[1], Laura Aldrovandi[1], Andrea Alessandrini[1,2], Paolo Facci[1]

[1] Centro S3, CNR-Istituto Nanoscienze, Via Campi 213/A, 41125 Modena, Italy
[2] Department of Physics, University of Modena and Reggio Emilia, Via Campi 213/A, 41125 Modena, Italy

ABSTRACT  We show that the activity of an ion channel is strictly related to the phase state of the lipid bilayer hosting the channel. By measuring unitary conductance, dwell times, and open probability of the K$^+$ channel KcsA as a function of temperature in lipid bilayers composed of POPE and POPG in different relative proportions, we obtain that all those properties show a trend inversion when the bilayer is in the transition region between the liquid disordered and the solid ordered phase. These data suggest that the physical properties of the lipid bilayer influence ion channel activity likely via a fine tuning of its conformations. In a more general interpretative framework, we suggest that other parameters such as pH, ionic strength, and the action of amphiphilic drugs can affect the physical behavior of the lipid bilayer in a fashion similar to temperature changes resulting in functional changes of transmembrane proteins.

## INTRODUCTION

The biological membrane configures a tremendously complicated ensemble which controls all the communications between the inner and outer regions of cells and organelles. It is mainly composed of lipids and proteins which cooperate to assure biological functions. In the recent years it has become always more evident that the lipid bilayer not only provides a medium for transmembrane protein folding and diffusion, but, with its dynamic heterogeneity, participates actively in the fine control of protein functionality (1,2). Exploiting model systems of continuously increasing complexity we are gaining knowledge on the unifying physical principles underlying lipid/transmembrane protein interactions (3). The mutual interplay between transmembrane proteins and lipids composing the bilayer can be the result of either specific interactions or non-specific collective bilayer properties (4). In the latter case, the lipid bilayer is considered as a continuum medium, characterized by certain mechanical properties (5). A transmembrane protein which accomplishes its function by a conformational change involving the lipid/protein interface will be influenced, as far as its functionality is concerned, by the lipid bilayer's properties (6,7). These include bilayer thickness, compressibility, lateral pressure profile, and surface charge. Many of these properties change abruptly during the bilayer's main phase transition which represents the switching of the bilayer organization from the solid ordered phase, characterized by lateral and molecular order, to the liquid disordered phase in which both levels of order are lost. For example, fluctuations in area, enthalpy and volume are enhanced and slowed down in the phase transition region, leading to a more compressible bilayer (8-11). It has already been reported that the Arrhenius plots of the functional activity of many membrane bound enzymes or transport membrane proteins show breaks which are considered hallmarks of lipid phase transitions (12). From a biological point of view, the proximity of a system to a phase transition assures a convenient and very efficient way to control biological functions with a very small variation in the environmental parameters. In fact, the transition can be induced by a change in temperature or in other physical-chemical parameters. This condition has been recently considered important in many biologically relevant situations, including dynamic heterogeneity in lipid bilayers and the cytoskeleton (13,14).

Ion channels represent important, paradigmatic examples of transmembrane proteins. The effect of temperature on the ion channel activity, including the unitary conductance and the gating properties, has been studied in the past to characterize the energetic requirements involved in the permeation process (15-18). Many reports have found non-linear Arrhenius plots of the channel activity both for the conductance and for rate constants (12). The discontinuity in the slope is usually interpreted as a strong indication of a phase transition, whereas, a continuous change of the slope is expected in cases of a broad two-phase coexistence region. However, even if the observed behavior of a channel can be understood in terms of membrane thermodynamics, there are no direct experimental evidences connecting directly the channel functional behavior to the physical state of the membrane at the single molecule level.

Here we used the ion channel KcsA to study the effect of the lipid bilayer phase state on protein ion channel functionality. KcsA from *Streptomyces Lividans* is a homotetrameric K$^+$ channel which can be gated by cytoplasmic pH changes (19). Its open probability increases at acidic pH with a half transition at pH 5.5. KcsA is the first K$^+$ channel whose structure has been solved by X-ray crystallography, shedding light on many of the structural properties of K$^+$ ion channels (20,21). Due to its well known structure, KcsA has become the experimental model system of choice to study ion channel and transmembrane protein properties also in relation to protein/lipid interaction. Generally, one assumes that the transition to the open state involves an expansion of the transmembrane helices against the lipid bilayer. Thus, protein conformational variations are likely to be affected by the physical state of the membrane (22).

## MATERIALS AND METHODS

### Sample preparation

The lipids 1-palmitoyl-2-oleoyl-*sn*-glycero-3-phosphoethanolamine (POPE) and 1-palmitoyl-2-oleoyl-*sn*-glycero-3-[phospho-*rac*-(1-glycerol)] (sodium salt) (POPG) were purchased from Avanti Polar Lipids (Alabaster, USA) and used without further purification. Stock solutions (in CHCl$_3$) were mixed to obtain the desired lipid molar ratios of 3:1, 1:1 and 0:1 (POPE:POPG). Then the chloroform was evaporated under a flow of nitrogen while heating the sample in a water bath at 50 °C. Thereafter, the sample was kept under vacuum (10$^{-2}$ mbar) for at least 4 hours to remove residual chloroform molecules. Samples which were not immediately used were flushed with nitrogen and stored at -80 °C. For the functional studies we added 100 μl of chloroform in order to obtain a final concentration of 5 mg/ml prior to usage. For protein reconstitution the lipids were rehydrated in a buffer solution of 450 mM KCl, 25 mM Hepes at a pH of 7 in order to obtain a final lipid concentration of 10 mg/ml. The sample was stirred at about 30 °C for 1 hour. During this time the sample was vortexed at least two times. For DSC measurements we added 125 mM KCl, 25mM KOH, 10 mM potassium dihydrogen citrate at pH values of 3, 4.5 and 6.5 in order to obtain a final concentration of 10 mM. The sample was stirred at about 30 °C for 1 hour. During this time the sample was vortexed at least two times.

### KcsA expression, purification and reconstitution

KcsA with an additional N-terminal hexahistidine sequence was expressed in BL21(DE3) cells grown in TB medium. Protein expression was induced by addition of 0.2 mg/l anhydrotetracycline (Acros Organics). The expressed protein was extracted with 20mM decylmaltoside (DM) and purified by Nickel affinity chromatography on a HisTrap FF crude column (Amersham Biosciences). The protein was eluted in 5mM DM, 25 mM KCl, 50 mM NaH$_2$PO$_4$, 32 mM NaOH and 400 mM Imidazole, pH 7.0. Immediately after Nickel affinity purification, the protein was concentrated and reconstituted by dialysis into POPE:POPG 3:1 (molar/molar) lipid vesicles following the procedure described in (23) with a protein-to-lipid ratio (w/w) ranging from 0.001 to 0.005 (functional studies) or 0.1 to 0.5 (atomic force microscopy). The prepared proteo-liposomes were stored at -80°C until usage. We performed the functional studies in varying ratios of POPG and POPE. Independently of this we always used the proteo-liposome preparation with POPE:POPG 3:1. It was shown before that the functional data depends solely on the lipid ratio used for the black lipid membrane formation (24).

### Single Channel Conductance Measurements

Vertical planar lipid bilayers were prepared over a round aperture with a diameter of 150 μm to 200 μm in a Teflon foil (25 μm thickness) separating two compartments. Prior to usage, the setup was cleaned with acetone, bidistilled water, and acetone and then it was dried under a nitrogen stream. After waiting 15 min the round aperture was prepainted using a solution of n-pentane/n-hexadecane (10:1). After 15 min the two compartments were filled with 1 ml of the desired buffer solution of 125mM KCl, 25mM KOH, 10 mM potassium dihydrogen citrate at pH 3 and pH 6.5 in the two compartments respectively. Then we added two Ag/AgCl electrodes which were embedded in agar-salt bridges (1.5 % agar, 3M KCl). The grounded electrode was inserted in the compartment with the buffer at pH 6.5 (cis side) and the input electrode was inserted to the one at pH 3 (trans side). Then we added 10-15 µl of the 5 mg/ml lipid in chloroform solution with the desired POPE to POPG molar ratios to both compartments. We waited for 15 min after which we raised the water levels above the aperture by adding another 1 ml of the respective buffer solutions to both compartments. At first we obtained a capacity in the order of 0.2 µF/cm$^2$ resulting from the formation of a hexadecane layer. This layer could not be destroyed even by applying holding potentials above 500 mV. We destroyed this layer mechanically and controlled the quality of the Ag/AgCl electrodes by determining their offset voltage. Electrodes requiring an offset-voltage of more than 10 mV were discarded. We then prepared a lipid bilayer across the aperture. In general we obtained lipid bilayers with a specific capacity of about 0.4 to 0.8 µF/cm$^2$. Applying voltages between 300 mV to 500 mV resulted in the breaking of the lipid bilayer. We then thawed the proteo-liposomes and sonicated them in an ultrasonic bath for 30 sec. Then we added 5 µl of proteo-liposome solution to the trans side (pH 3). For temperature-control of the planar bilayer setup we used water circulation around the bilayer cell. The temperature was determined in a separate gauching measurement as a function of the heat bath temperature. The temperature was measured by a digital thermometer Fluke 16 (Fluke, Italy) equipped with a small K-thermocouple probe (Thermocoax GmbH, Germany). Temperature was changed at a rate of 5 °C/h as done in the Differential Scanning Calorimetry experiments (see below). During the recording of ion current traces temperature did not vary significantly (less than 0.1 °C). The planar bilayers can be easily destroyed at relatively high holding potentials and become more and more sensitive in the phase transition regime where we had to limit the holding potentials between -100 mV and 100 mV. For data acquisition we used an EPC 8 patch clamp amplifier (Heka, Germany). The data were recorded using GePulse software (Pusch lab, CNR-Institute of Biophysics, Genova, Italy) at a rate of 10 kHz and they were subjected to a 7-pole Bessel filter at 1kHz. Depending on the purpose of the measurement we recorded traces with a length of time ranging from 5 min to 20 min. Data were analysed without further filtering using either Analysis software (Pusch lab, CNR-Institute of Biophysics, Genova, Italy) or QUB software (SUNY at Buffalo, Buffalo, USA). Ion current histograms were fitted with Gaussian distributions. For an analysis of the ion channel's kinetics only sequences which showed the presence of only one active channel were considered. Then, the ion current traces were idealized in noise-free open and closed transitions using the segmental k-means (SKM) algorithm (25,26). The outcome was controlled visually. Dwell-time data were assembled in a logarithmic time axis on the abscissa and a square root ordinate which was normalized for representation. The non-normalized dwell-time histograms were fitted with exponential components.

**Differential Scanning Calorimetry**

Differential Scanning Calorimetry (DSC) measurements were performed with a VP-DSC from Microcal (Northhampton/MA) using high feedback mode at a scan rate of 5 °C/h. Suspensions of POPE:POPG 3:1 or 1:1 in buffer solutions of 125mM KCl, 25mM KOH, 10 mM potassium dihydrogen citrate at pH values of 3, 4.5 and 6.5 at a lipid concentrations of 10mg/ml were sonicated in an ultrasonic bath for 30 seconds in order to obtain small unilamellar vesicles prior to the measurements. Excess heat capacity profiles of the sample were directly measured in the calorimeter cell.

# Results

### Effect of temperature on KcsA conductance

In this study we aimed at investigating the functional properties of an ion channel as a function of the lipid bilayer's physical state. This aspect required a detailed understanding of the physical properties of the lipid system under varying conditions. Thus, at first we performed Differential Scanning Calorimetry (DSC) characterizations of Small Unilamellar Vesicles (SUVs) composed by POPE:POPG 3:1 and 1:1 at different pH. Fig.. 1a displays the excess heat capacity profiles as a function of pH for a mixture of POPE:POPG 3:1 measured from low to high temperature. For all the displayed pH values the melting transition regime lies between 14 °C and 26 °C and shifts to higher temperatures for acidic pH. Fig. 1b reports the transition midpoint temperature ($T_m$) *vs* pH for both the upscan (low temperature to high temperature) and the downscan (high temperature to low temperature) directions. It appears that $T_m$ depends linearly on pH. The melting transition shows hysteresis so that the melting profiles are shifted by about 1.5 °C to lower temperatures when cooling the bilayer with respect to the heating scans. We have also compared the phase transition regime of sonicated and non-sonicated liposomes of POPE:POPG 3:1. We have not found any appreciable difference between these systems. This investigation allowed us to establish that for the lipid system at issue the phase transition from liquid disordered to solid ordered phase occurs around room temperature (from 18° to 23°C). In the case of the pure POPG bilayer the transition occurs at a temperature below 0°C at neutral pH (27). The phase transition in the presence of both POPE and POPG is characterized by a broad phase coexistence region where the liquid disordered phase will be enriched in POPG and the solid ordered phase will be enriched in POPE. Hence, many of the reported experiments on ion channel functionality performed on Black Lipid Membranes (BLM) with this lipid composition are obtained in a temperature range near the phase coexistence state.

After having established the thermodynamics of our lipid mixtures we performed temperature-dependent single channel conductance measurements of KcsA incorporated in varying molar mixtures of POPE:POPG (3:1, 1:1, 0:1). Traces of the KcsA unitary current as a function of temperature in an asymmetric pH configuration using a virtually solvent-free POPE:POPG 3:1 lipid bilayer are reported in Fig. 2a. Fig. 2b shows KcsA conductivity at two holding potentials (50 mV and 75 mV) in a larger temperature range. The two lines represent linear fits to the conductivity $vs$ temperature data in the higher temperature monotonic region of the graph (from 22 °C to 28 °C). From our data, it follows that the conductance of the channel is non-monotonic with temperature. Indeed, upon cooling, at a temperature of 22°C, an increase of the conductance is observed. Upon further decreasing the temperature, a maximum in the conductance is reached followed by a further decrease. The same behavior is observed for both holding potentials. For a POPE:POPG 1:1 lipid composition, KcsA conductance initially decreases, but, at 19°C it starts to increase. Interestingly, a monotonic decrease is observed all over the investigated temperature range for a pure POPG bilayer (Fig. 1c). Fig. 1c also highlights how KcsA conductivity depends on the lipid composition of the bilayer; specifically it increases with the POPG fraction (24). To further analyze the anomalous behavior of the conductance we report the channel conductance $vs$ temperature, after subtracting the linear trend observed for temperatures above 22°C and 19°C for the 3:1 and 1:1 mixtures, respectively (Fig. 1d). Comparing these trends with the corresponding excess heat capacities (both systematically shifted by -1.5 °C) we found a surprising correlation between the variations in heat capacity and ion conductance for both lipid mixtures. Comparing the two mixtures, POPE:POPG 3:1 and 1:1, one has to point out that the anomalous behavior in the ion channel conductance is modulated by the lipid composition used in accordance to the expected changes in the phase transition temperature. Most important, the linear relationship between KcsA single channel conductance and temperature in a pure POPG bilayer is consistent with the absence of a lipid phase transition in the investigated temperature range. In the case of POPE:POPG 3:1, KcsA conductance, after reaching a maximum, starts to decrease upon reducing temperature. Instabilities, due to the growth of solid ordered domains, prevented us from performing measurements at temperatures much lower than that at which ion conductance reached its maximum.

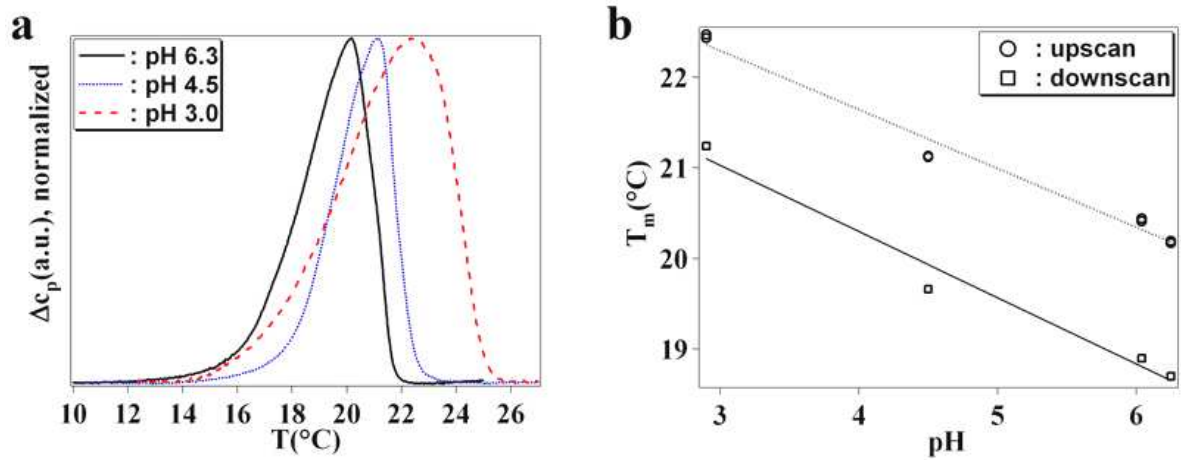

**Figure 1.** Thermodynamic characterization of the main phase transition behavior of POPE:POPG 3:1 bilayers. (a) DSC curves are reported for the POPE:POPG 3:1 at varying pH values. (b) Variation of the transition midpoint temperature of the POPE:POPG 3:1 bilayers as a function of pH. Both the upscan and the downscan of the DSC measurements are reported highlighting the hysteretic behavior.

Many physical properties of the lipid bilayer change abruptly in the proximity of the main phase transition. These also include the surface charge density in a lipid bilayer containing charged lipid molecules. The surface charge density increases in the solid ordered phase due to the reduced area per molecule. This charge density may in principle have a role in the enhanced channel conductance since it can alter the local effective concentration of $K^+$ near the mouth of the channel. The effect of the lipid bilayer surface potential on channel conductance has been thoroughly studied (28-30). In general, the lipid bilayer surface potential is screened by ions in the solution so as to result strongly attenuated at the mouth of the ion conducting pore protruding off the lipid bilayer surface. For example, the Debye length, which characterizes the exponential decay of the potential in solution, in 100 mM KCl is 9.6 Å, meaning that the potential at the mouth of the pore for KcsA would be only 6% of the lipid surface potential (30). Marius et al.(24) found a small effect of the surface potential on the channel behavior. For example, the mechanical properties of a bilayer are affected by ionic strength due to the possible influences on the phospholipid headgroup interactions (31-32). Probably, here a concurrent effect of different mechanisms could be responsible for the channel conductance behavior.

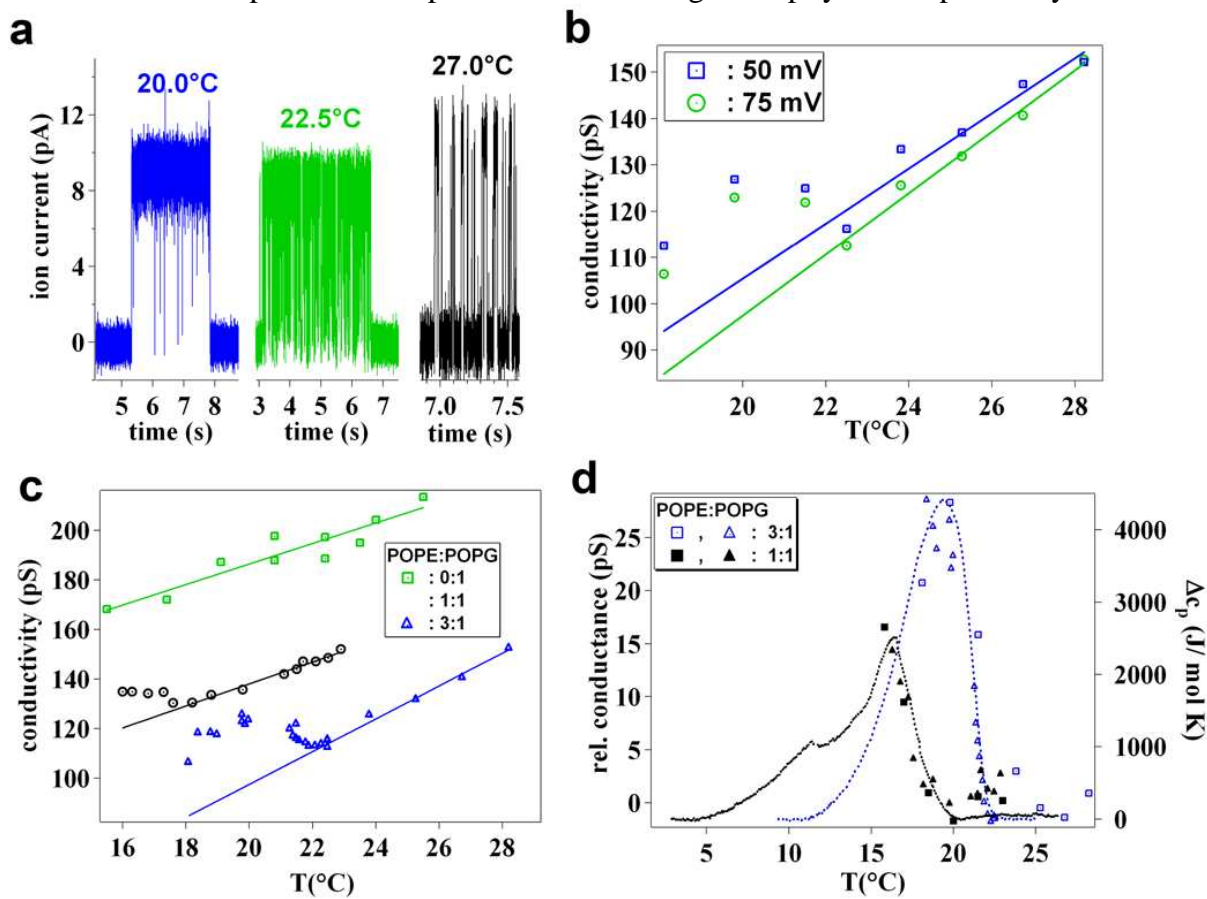

**Figure 2.** Effect of temperature on KcsA conductance. (a) Examples of traces of the current for KcsA in POPE:POPG 3:1 at three different temperatures: 27.0°C, 22.5°C and 20.0°C. Holding potential: 75 mV. (b) Plot of the conductivity of KcsA reconstituted in POPE:POPG 3:1 as a function of temperature for two holding potentials (□: 50 mV and ○: 75 mV). The two lines are the linear fits to the experimental data in the regions of monotonic behavior of the channel conductance (from 22.5 °C to 28.5 °C; ---: 50 mV; —: 75 mV). Ion conductance changes linearly in a temperature region between 28.5 °C and 22.5 °C, but upon further cooling, at a temperature of about 22.0 °C a trend inversion appears. (c) Temperature-dependent ion conductance of KcsA in BLMs of different POPE:POPG compositions at a holding potential of 75mV. The solid lines are the respective linear fits in the high temperature region. A non-monotonic behavior is observed for the two POPE:POPG mixtures (○: 3:1; Δ: 1:1), but a pure linear trend is found for POPG (□). The trend inversion happened at a lower temperature when KcsA was reconstituted in POPE:POPG 1:1 than in 3:1, accordingly to the shift of the phase transition region for the two lipid compositions. (d) Superposition of the DSC traces (shifted by -1.5 °C) for the POPE:POPG 3:1 (—) and 1:1 (---) lipid bilayers with channel conductance at a holding potential of 75mV after subtraction of the linear fits obtained outside the lipid phase transition. The trend was extended into the transition region. Data points were taken at constant temperature (□: 3:1; ■: 1:1) or during the temperature change (Δ: 3:1, ▲: 1:1). The changes in ion conductivity take place proportionally to the variation of the system's specific heat capacity.

It has been demonstrated that by changing the acyl chain length of the bilayer hosting the KcsA channel, the tilt angles of the transmembrane helices of KcsA vary accordingly to the hydrophobic thickness of the lipid bilayer (33). This evidence outlines the elasticity of the channel in the lipid bilayer, suggesting that the structural changes of the protein could control its conductance. In such a framework, the gating transition differences between KcsA conductance in different bilayer compositions. Furthermore, it has been demonstrated that surface active agents are able to affect the unitary ion channel conductance without involving an electrostatic variation of the bilayer (22). Thus, even if the electrostatic effect may have a role, it does not rule out a possible effect of the lipid bilayer physical properties on the involves a variation of the channel conformation at the lipid/protein interface. Thus, the mechanical properties of the lipid bilayer could exert a sort of fine-tuning of the protein conformation which could alter the free energy variation for an ion passing the filter. Furthermore, small variations in the radius of the KcsA inner pore can produce drastic changes in channel conductance (34).

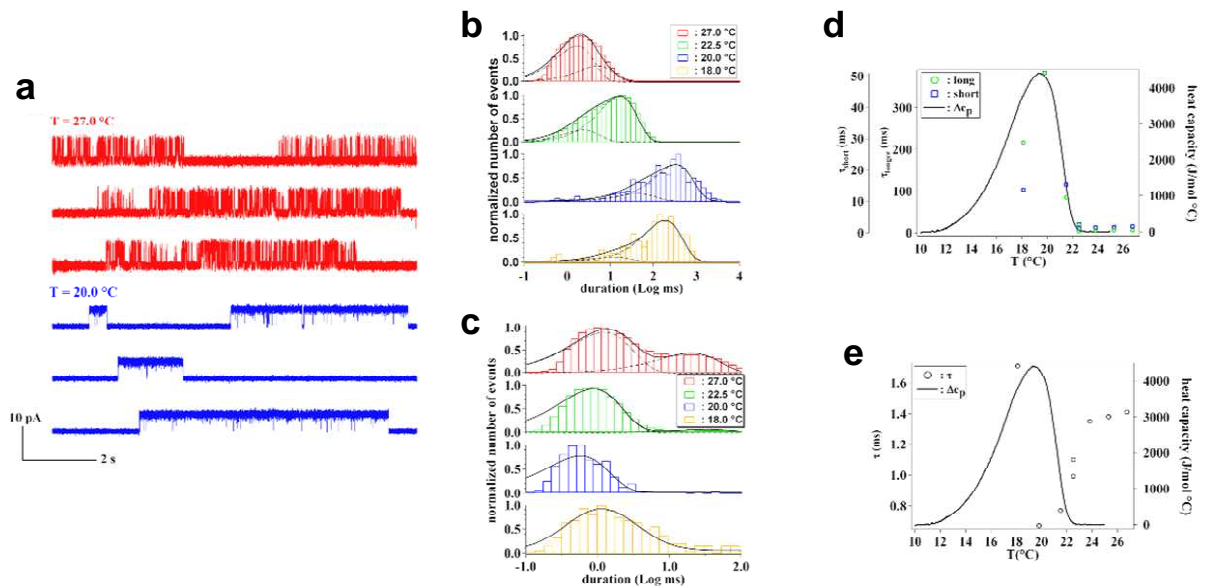

**Figure 3**. KcsA dwell times as a function of temperature at a holding potential of 50mV. (a) Examples of ion current traces at temperatures of 27.0 °C and 22.0 °C in a POPE:POPG 3:1 lipid mixture. Channels are open for longer times at the temperature of 20.0 °C where the ion conductance presents a maximum. (b)-(c) Dwell-time histograms for the open (b) and closed (c) states, for KcsA reconstituted into planar bilayers of POPE:POPG 3:1. The open dwell-time and closed dwell-time distributions were fitted with exponential components (·····) and the overall fit (—). (c-d) Variation of the characteristic open (longer time: ○, slower time: □) and closed times (○) with temperature compared to the DSC trace (—) of the corresponding lipid bilayer mixture. Open times show the same trend as the specific heat capacity. Closed times present a minimum at the temperature corresponding to the maximum of the excess heat capacity and open times.

However, also specific interactions between KcsA and anionic lipids such as POPG have been considered (24). In order to better investigate the reasons for the observed conductance variation, we studied the kinetic features of the channel activity as a function of temperature.

**Effect of temperature on KcsA dwell times**

KcsA activity is modulated by pH variations at the intracellular compartment, where an acidic pH increases channel open probability (35-39). Upon opening, KcsA undergoes a further conformational change in which a second gate inactivates the channel (37,40,41). The molecular determinants of both pH sensitivity and inactivation have been elucidated. In steady state measurements, the dominant gating mechanism is usually attributed to the transition from an inactivated state to an open and conductive state. Consequently, the low open probability of KcsA has been interpreted on the basis of the recovery mechanism from inactivation (40). Another important aspect of KcsA gating is the presence of a weak voltage dependence even if the channel does not exhibit an evident voltage-sensing domain (42). A typical trace of single-channel activity is represented by bursts of activity separated by long closure periods. Intraburst activity is characterized by fast fluctuations of the channel between a conductive and non-conductive state. It has been shown that intraburst activity can involve different modes of activity even under apparently identical experimental conditions. In fact, the kinetic behavior of KcsA presents a combination of three distinct modes of channels activity, the low opening probability, high opening probability, and the flickering mode, which appear randomly from patch to patch (43).

In Fig. 3a we show representative ion current traces of KcsA in POPE:POPG 3:1 at the two temperatures of 27.0 °C and 20.0 °C. At a temperature of 20 °C we found a maximum for the KcsA conductivity in this lipid mixture (see Fig. 2d). Comparing the traces in Fig. 3a it becomes immediately clear that the KcsA behavior is affected by the temperature as far as characteristic times are considered. We calculated the life time distributions at varying temperatures for both the open and closed states. The results are reported for four different temperatures in Fig.s 3b and c. The open time distributions, considering exponential components, show always two characteristic times which we define as the longer and shorter opening time. In Fig. 3d the longer and shorter opening times are reported as a function of temperature along with the DSC trace for the same lipid mixture used in the BLM experiment. While both times do not change appreciably with temperature above the lipid phase transition region, once the lipid bilayer enters the phase transition, both times increase attaining a maximum along with the excess heat capacity of the bilayer and then decrease upon further lowering the temperature. In the phase transition region the open time distribution shifts towards longer durations (44). The presence of two characteristic open times can be explained by a switching of the channel between different burst modes or by the presence of more than one channel (43). Independently of the interpretation, both times follow the same trend of the excess heat capacity of the lipid bilayer phase transition. The closed time distribution for KcsA (Fig. 3c) is described by one characteristic time which decreases at the lipid phase transition (Fig. 3e) and increases again once the maximum in the excess heat capacity is reached. The open times are more affected by temperature than the closed ones. The longest time observed in the distribution of the closed times can be ascribed to the interburst interval. The same general trend with a slowing down in the transition regime that reaches a minimum at the transition midpoint was obtained using the POPE:POPG 1:1 mixture.

**Effect of temperature on KcsA normalized open probability**

The third functional parameter which is revealed by traces of single channel conductance is the normalized open probability. Figs. 4a and b show the normalized open channel probability *vs* temperature in the case of POPE:POPG 1:1 and POPE:POPG 3:1 lipid mixtures, assuming a constant number of active channels. At a temperature corresponding to the onset of the lipid phase transition for the used lipid mixture, also the normalized open probability shows a trend inversion. Initially it decreases when lowering the temperature in the region above the lipid phase transition. At 19°C, which corresponds to the onset of the phase coexistence region for the 1:1 lipid mixture, it starts increasing until it reaches a maximum and then decreases again. Most of the effect on the normalized open probability is due to a decrease in the closing rates, as it can be deduced by the temperature dependence of the open and closed dwell times. The behavior of open and closed times supports the interpretation of the obtained results in terms of a variation of the normalized open probability rather than a variation in the number of active channels. Fig. 4b shows the normalized open probability for the POPE:POPG 3:1 mixture. In this case, the trend inversion for the open probability occurs at a temperature of 22°C, along with entering the lipid phase transition. The observed behavior for the normalized open probability in both cases points again to a strong connection between the thermodynamics of the lipid membrane and the functioning of the channel.

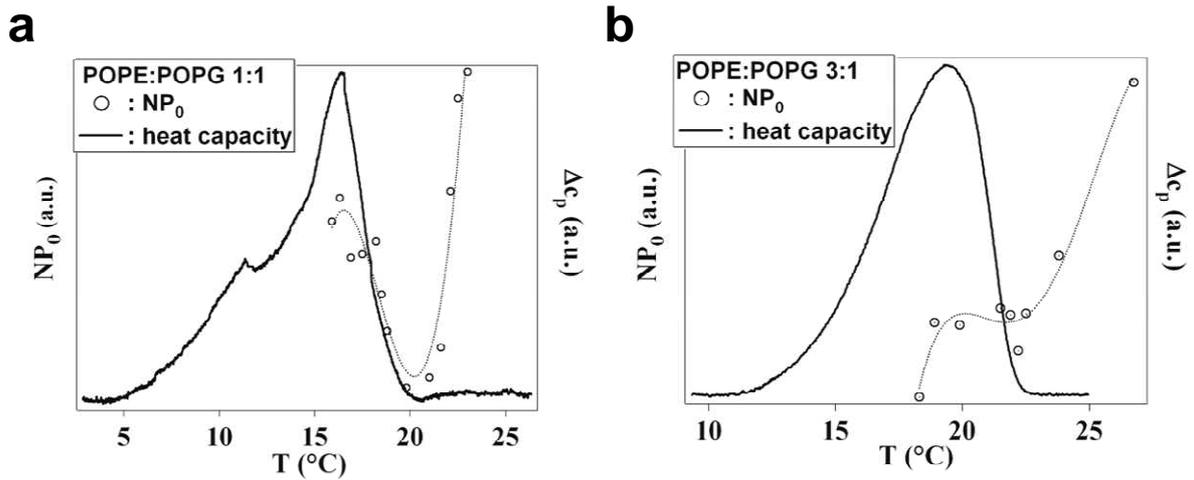

**Figure 4.** Normalized open probability for KcsA in lipid bilayers of different composition compared with the respective DSC analysis. (a) POPE:POPG 1:1; (b) POPE:POPG 3:1. The normalized open probability NP$_0$ (○) in arbitrary units referring to the right ordinate and the specific heat capacity (—) in arbitrary units referring to the left ordinate are reported. In both cases the open probability has a trend inversion when the system enters the phase transition region.

**DISCUSSION**

Variations in temperature can affect many biological processes. In the case of proteins in lipid bilayers, temperature can affect both the properties of the lipid ensemble and the conformational transitions of the proteins which underlie their activity. There are several well documented examples of how the activity of membrane bound proteins can be modulated by factors affecting the physical properties of the lipid bilayer. These factors include the lipid composition, the lateral pressure profile, the temperature, the curvature and the thickness of the bilayer (29,45,46). Lipid composition can affect ion channel conductance in different ways: causing specific interactions between lipids and ion channels, altering the surface charge of the lipid bilayer, and modifying lipid bilayer physical properties. The physical properties of the lipid bilayer can affect the energetic cost of protein conformational changes and can perform a fine tuning of protein conformation. In the particular case of ion channels, different protein conformations may be associated with rate-determining barriers to ion movements influencing the observed conductance of the channels. Activity of KcsA is strictly connected with its inactivation state. Indeed, KcsA is thought to reside mainly in a long-lived inactivated state with an overall low open probability. The E71A mutant is able to prevent the transition to this inactivated state by removing the Glu71-Asp80 interaction. The intraburst activity is related to the equilibrium between the open-conductive state and an inactivated state. An effect of the lipid phase transition on the intraburst activity where the transition to the inactivated state involves structural rearrangements in the selectivity filter is not immediately intuitive. In fact this conformational transition seems not to involve the protein/lipid boundary. However, the variation of the channel unitary conductance in the phase transition region supports the idea that the conformation of open KcsA in this region is different from the conformation outside the transition. It is possible that the new conformation alters the rate of transition to/from the inactivated state.

It should be noted that by inducing a phase transition in a POPE:POPG supported lipid bilayer containing KcsA molecules, these partition preferentially in the vicinity of the domain boundaries and in the liquid disordered phase of the bilayer (46) (see also Supplementary Material). The domain boundaries are characterized by strong local fluctuations (47) and it is expected from thermodynamic considerations that the liquid disordered regions are richer in POPG with respect to the solid ordered ones. Thus, KcsA may experience an effective increase in POPG concentration. This phenomenon could also influence the channel conductance in the lipid bilayer via specific lipid/protein interactions. Indeed, it has already been demonstrated that an increase of the POPG content of lipid bilayers induces an increase in KcsA conductance and open probability (24) (see Fig. 2c). The sorting of membrane proteins to specific lipid domains with a consequent modification of their function recalls the case of lipid rafts, where the presence of specific lipids promotes the formation of a liquid ordered phase enriched in cholesterol and sphingolipids. However, two considerations are in favor of an interpretation of the observed variation in functionality not based exclusively on a lipid specific effect. The first consideration refers to the observed decrease of the KcsA conductance after having reached a maximum inside the transition region. This decrease occurs while the POPG enrichment of the liquid disordered domains is still going on. The second consideration is connected to a quantitative analysis of the dwell times. We observed a 40-50 fold increase of the dwell times in the POPE:POPG mixtures when the system entered the lipid phase transition region. The longest opening time at a holding potential of 50 mV was 380 ms. This value is significantly higher than the dwell times determined for KcsA in a pure POPG bilayer at the same temperature. These findings can be well explained considering the changes in the lipid bilayer's physical state in the vicinity of the phase transition. There, the bilayer reaches a maximum compressibility along with heat capacity (8,32), thus facilitating the protein's conformational changes. In addition, it should be mentioned that the associated time scales of the bilayer's physical properties are proportional to the bilayer's excess heat capacity (9,10).

## CONCLUSION

In conclusion, we have reported that the activity (conductance, dwell times, normalized open probability) of KcsA is finely tuned by the physical state of the lipid bilayer. Here, we studied the liquid disordered to solid ordered transition, but it should be possible to extend the same physical concepts to other lipid transitions involving different phases. These experiments prompt further attention to the regulation of membrane protein activity by the lipid environment. Indeed, many of the temperature induced variations in a lipid bilayer can be equally induced by changes in other parameters such as lipid composition (i.e. lipid synthesis), ionic strength, pH or by the presence of amphiphilic drugs (22).

## ACKNOWLEDGMENTS

KcsA clones were a kind gift of C. Miller. The authors thank C. Miller and A.G. Lee for their critical reading and very useful comments on the manuscript. The authors are indebted to M.F. Schneider and C. Westerhausen for performing DSC measurements on POPE:POPG 1:1 and to ML. Caiazzo for assistance in protein expression, purification, and reconstitution. Partial financial support by Italian MIUR FIRB project Italnanonet is acknowledged. The authors declare they have no competing interests.

**Supporting Material**

# Changes in single K[+] channel behavior through the lipid phase transition

Heiko M. Seeger[1], Laura Aldrovandi[1], Andrea Alessandrini[1,2], Paolo Facci[1]
[1] Centro S3, CNR-Istituto Nanoscienze, Via Campi 213/A, 41125 Modena, Italy
[2] Department of Physics, University of Modena and Reggio Emilia, Via Campi 213/A, 41125 Modena, Italy

## Supporting Methods

### Supported Lipid Bilayer Preparation

Supported lipid bilayers were prepared by the vesicle fusion technique. The protein-lipid suspension was sonicated for 30s in an ultrasonic bath to obtain small unilamellar proteo-liposomes (SUVs). Then we added 70 µl of our proteoliposome suspension on a freshly cleaved piece of mica. The lipid suspension was incubated for 15 min at 32°C and then rinsed abundantly with 450 mM KCl, 25 mM Hepes, pH 7 buffer solution. The solution was then exchanged for the imaging solution (150mM KCl, 10mM potassium dihydrogen citrate at pH 7) by extensive rinsing. Then the mica support with the formed lipid bilayer was mounted on the temperature-controlled stage of the AFM.

### Atomic Force Microscopy

Atomic Force Microscopy (AFM) experiments on supported lipid bilayers with reconstituted KcsA proteins were performed with a Bioscope equipped with a Nanoscope IIIA controller (Veeco Metrology, USA).

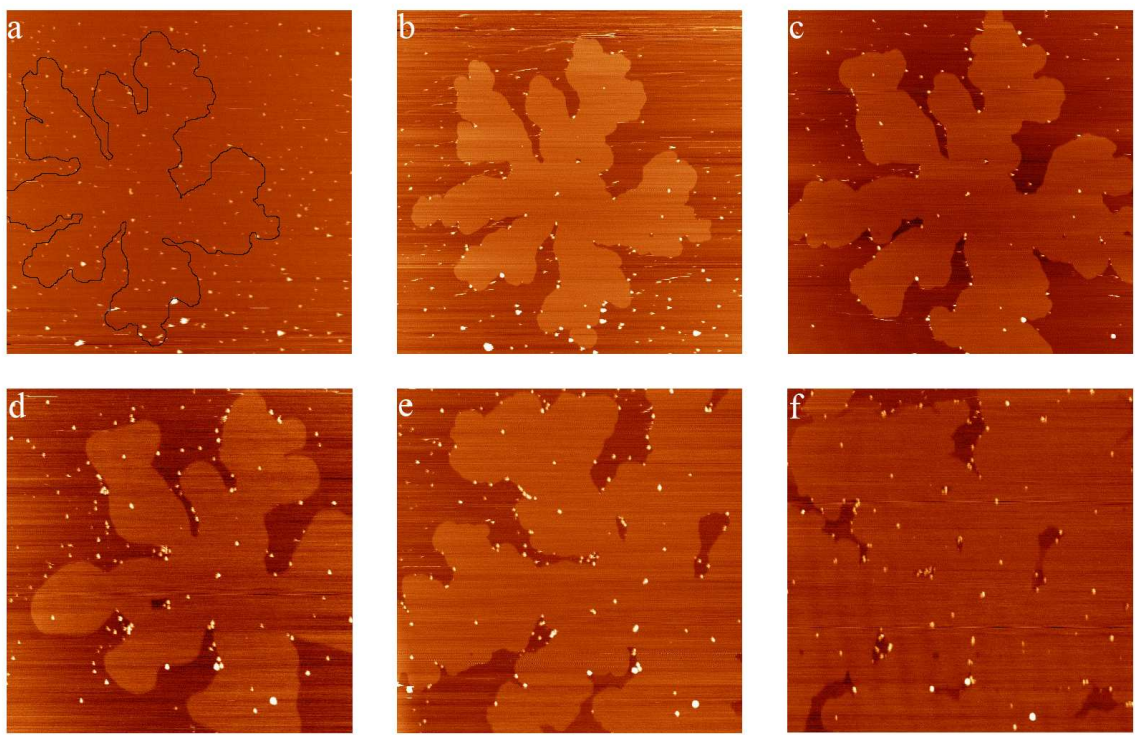

Figure S1. Series of AFM images (10 µm x 10 µm) of the redistribution behaviour of KcsA reconstituted in a SLB of POPE:POPG 3:1. (a) The SLB was equilibrated in the liquid disordered phase. The KcsA proteins are randomly distributed in the SLB at 28 °C. The black line is the outline of the solid ordered domain which formed in b. (b) A solid ordered domain (lighter area) was induced by cooling at 26.5 °C. The KcsA proteins were excluded from the solid ordered region. (c-d) The solid ordered domain was allowed to equilibrate at 26.5 °C. (e-f) The phase transition and the protein redistribution proceed upon further temperature decrease to (e) 25.0 °C and (f) 23.0 °C.

Imaging was performed in tapping mode at a scan rate of 1-2 lines/s using triangular silicon nitride cantilevers (Olympus OMCL-TR400PB-1, Japan) with a nominal spring constant of 0.09 N/m and a resonance frequency in liquid between 8-9 kHz. The force applied to the membrane was adjusted to the lowest possible value allowing reproducible imaging. The sample temperature was continuously monitored by a digital thermometer Fluke 16 (Fluke, Italy) equipped with a small K-thermocouple probe (Thermocoax GmbH, Germany) in direct contact with the imaging buffer.

# Supporting Results

In Figs. 2d., 3c, d and 4 of the manuscript we compared functional parameters with the excess heat capacity profiles for both the POPE:POPG 3:1 and POPE:POPG 1:1 lipid mixtures. In all cases the excess heat capacity profiles had to be corrected by -1.5 °C. This can be explained by the geometrical differences between the planar bilayers (functional studies) and the liposomes (DSC) and a possible influence of hexadecane residues in the planar lipid bilayer. It should be stressed that the shift was systematic for both mixtures -1.5°C. In general, we have followed changes in the bilayer's capacitance during temperature changes in order to estimate the BLMs transition behavior (1). We could conclude that both systems display transition regimes in a similar temperature regime, with a shift of -1.5 °C being feasible. It should be pointed out that in proximity of the melting phase transition of lipid bilayers many physical parameters change. These changes involve an enhancement of macroscopic fluctuations in enthalpy, volume and area (2). The fluctuation-dissipation theorem relates these changes to the heat capacity profile, meaning that the higher the excess heat capacity at a certain temperature the stronger are the fluctuations. In addition fluctuations slow down in the phase transition region (3,4). This means that all physical bilayer properties which are related to the system's fluctuations display the same kinetics and strength. One of these parameters is the bilayer's compressibility. The lipid bilayer becomes softer in the phase transition regime, reaching a maximum at the temperature corresponding to the highest excess heat capacity. It is very likely that a protein's conformational changes involving the lipid/protein interface are fine-tuned by these physical properties.

**Atomic Force Microscopy Experiments: Protein Redistribution in Dependence of Domain Coexistence**

In a recent study we studied the redistribution behavior of KcsA protein reconstituted in supported lipid bilayers (SLBs) after the evolution of a solid ordered domain in a previously liquid disordered bilayer (5). In Fig. S1 a series of AFM images of size 10 μm x 10 μm taken at different temperatures are shown. In Fig. S1a the SLB was in the liquid disordered phase. The protrusions out of the bilayer can be associated to the protein's cytoplasmic domain which was mainly exposed towards the AFM tip. The black line gives the outline of the solid ordered domain (lighter color) which formed upon cooling as shown in Fig. S1 b. The solid ordered domain is about 1.4 nm higher than the liquid disordered region. Comparing Fig. S2 a and S1 b it is evident that before the evolution of the solid ordered domain the protein's were randomly distributed, but, upon the formation of the solid ordered domain, the protein's were mainly excluded from the solid ordered region. Following the evolution of the domain in time (Fig. S1 b-d) and temperature (Figs. S1 e and S1f) it becomes evident that the proteins have the tendency to align along the domain boundaries or eventually start to cluster in the liquid disordered environment. It should be noted that the observed behavior is in agreement with the hydrophobic matching principle (5). We have also performed heating experiments. In these experiments the SLB was rapidly cooled at the beginning of the experiment, a solid ordered domain was present and then we heated the sample again. Representative zooms of two regions of the same SLB are given in Fig. S2. In both regions it is possible to follow the melting of the solid ordered domain. At the same time the proteins align along the domain boundaries and in some cases they follow the retracting interface between lipid domains. This can be attributed to an attractive potential towards the domain interface acting on the protein (6).

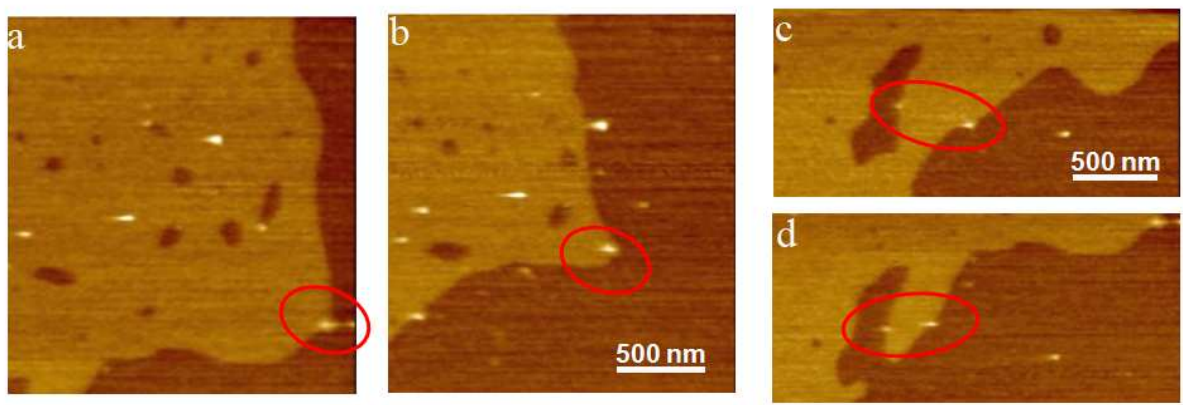

Figure S2. Redistribution of KcsA upon domain melting. (a) At 18.5 °C a KcsA protein (in the red ellipse) is aligned at the domain boundaries between the solid ordered and liquid disordered phase. (b) Upon heating to 19.5 °C the solid ordered domain melts and hence occupies a smaller area fraction. The protein stays aligned between the solid ordered/ liquid disordered interface. (c-d). The redistribution behaviour of two KcsA proteins were followed in another area of the SLB and at other temperatures of (c) 16.5 °C and (d) 18.0 °C.